\documentclass[twocolumn,nofootinbib,amsmath,amssymb]{revtex4}

\usepackage{graphicx}% Include figure files
\usepackage{epsfig}
\usepackage{dcolumn}% Align table columns on decimal point
\usepackage{bm}
\newcommand{\etal}{{\it et al.}}
\begin{document}

\title{Leptonic and Semileptonic Charm Decays from CLEO-c}% Force line breaks with \\

\author{S. Stone}
 \email{stone@physics.syr.edu}
\affiliation{Physics Department of Syracuse University\\
Syracuse NY, 13244, USA\\
}%

\begin{abstract}
I describe CLEO-c purely leptonic decay results leading to
$f_{D^+}=(222.6\pm 16.7^{+2.8}_{-3.4})~{\rm MeV}$,
$f_{D_s^+}=(280.1\pm 11.6 \pm 6.0) {~\rm MeV}$, and
$f_{D_s^+}/f_{D^+}=1.26\pm 0.11\pm 0.03$. Form-factor measurements
in Cabibbo favored and suppressed pseudoscalar decays are
presented. Some comparisons are made with theoretical predictions.
\end{abstract}
\maketitle

\section{Introduction}
Threshold production of $D^0\overline{D}^0$ and $D^+D^-$ mesons at
3770 MeV, and $D_s^+D_s^{*-}$ + $D_s^{*+}D_s^{-}$ mesons at 4170 MeV
in $e^+e^-$ annihilations have allowed CLEO-c to make precision
measurements using purely leptonic and semileptonic charm meson
decays.

\section{Purely Leptonic Decays}

To extract precise information from $B$ mixing measurements the
ratio of ``leptonic decay constants," $f_i$ for $B_d$ and $B_s$
mesons must be well known \cite{formula-mix}. Indeed, the recent
measurement of $B_s^0$ mixing by CDF \cite{CDF} has pointed out the
urgent need for precise numbers. The $f_i$ have been calculated
theoretically. The most promising of these calculations are based on
lattice-gauge theory that include the light quark loops
\cite{Davies}. In order to ensure that these theories can adequately
predict $f_{B_s}/f_{B_d}$ it is critical to check the analogous
ratio from charm decays $f_{D^+_s}/f_{D^+}$. Here I present the most
precise measurements to date of $f_{D_s^+}$, $f_{D^+}$
\cite{our-fDp,DptomunPRD} and $f_{D_s^+}/f_{D^+}$.

\begin{figure}[htb]
 \vskip -.4cm
 \includegraphics[width=0.3\textwidth]{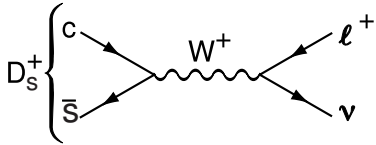}
 %\centerline{ \epsfxsize=2.2in \epsffile{Dstomunu.eps} }
 \vskip -.2cm
 \caption{\label{Dstomunu}The decay diagram for $D_s^+\to \ell^+\nu$.}
 \end{figure}

In the Standard Model (SM) the $D^+_{(s)}$ meson decays purely
leptonically as shown in Fig.~\ref{Dstomunu}. The decay width is
given by \cite{Formula1}
\begin{eqnarray}\label{eq:main}
\Gamma(D_{(s)}^+\to \ell^+\nu) &=& {{G_F^2}\over
8\pi}f_{D_{(s)}^+}^2m_{\ell^+}^2M_{D_{(s)}^+}\\
[4pt] &&\times\left(1-{m_{\ell^+}^2/ M_{D_{(s)}^+}^2}\right)^2
\left|V_{cq}\right|^2~, \nonumber
\end{eqnarray}
where $m_{\ell^+}$ and $M_{D_{(s)}^+}$ are the $\ell^+$ and
$D_{(s)}^+$ masses,  $|V_{cq}|$ is the CKM element appropriate to
either $D^+$ $(V_{cd})$ or $D_s^+$ $(V_{cs})$ decay and $G_F$ is the
Fermi constant.

 New physics can affect the expected widths; any
undiscovered charged bosons would interfere with the SM $W^+$. These
effects may be difficult to ascertain, since they would simply
change the value of the $f_i$'s. The ratio $f_{D_s^+}/f_{D^+}$,
however, is much better predicted in the SM than the values
individually. Akeroyd predicts that the presence of a charged Higgs
boson would suppress this ratio significantly \cite{Akeroyd:2003}.
In addition, the ratio of decay rates to different leptons are fixed
only by well-known masses in Eq.~\ref{eq:main}. For example, the SM
prediction for $\Gamma(D_s^+\to\tau^+\nu)/ \Gamma(D_s^+\to\mu^+\nu)$
is 9.72. In general, any deviation from a predicted ratio would be a
manifestation of physics beyond the SM, and would be a clear
violation of lepton universality \cite{Hewett}.

\begin{figure}[htbp]
 \vskip -.2cm
\includegraphics[width=0.5\textwidth]{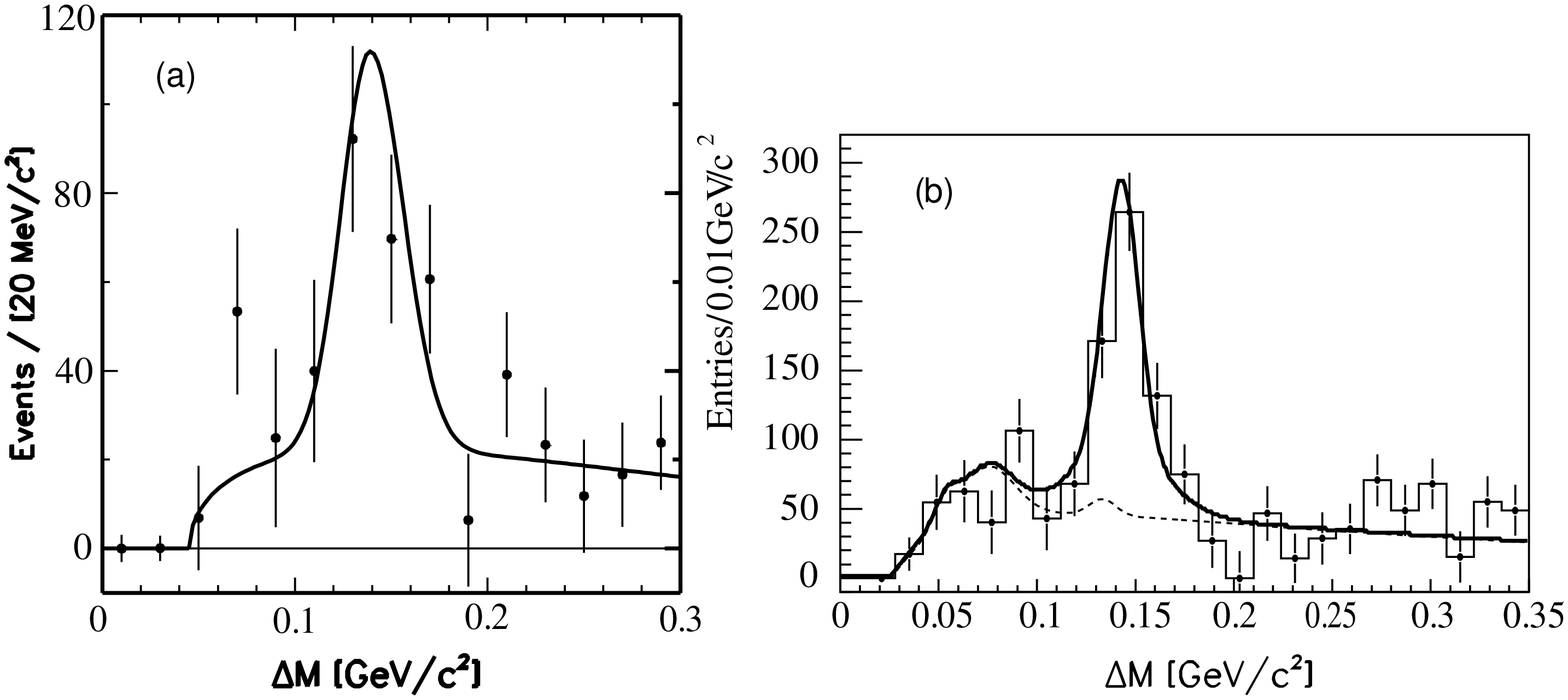}
\vskip -.2cm
 \caption{\label{DM} The $\Delta M$ distributions for
$\mu^+$ candidates after the $e^+$ subtraction from (a) CLEO and (b)
BaBar. The solid curves are fits to signal plus background.}
 \end{figure}

CLEO previously measured $f_{D_s^+}$ using 4.8 fb$^{-1}$ of
continuum annihilation data at or just below the $\Upsilon$(4S)
\cite{Chadha}. This analysis introduced a number of new ideas: (i)
The $\gamma$ and $\mu^+$ from $D_s^{*+}\to\gamma D_s^+;$ $D_s^+\to
\mu^+\nu$ were detected directly, and the $\nu$ 4-vector was
inferred from missing energy and momentum measurement in half of
the event, where the event half was determined using the normal to
the thrust axis. (ii) The $\nu$ 4-vector was corrected to get the
right $D_s^+$ mass and $\Delta M=M(\gamma\mu^+\nu)-M(\mu^+\nu)$
was examined (see Fig.~\ref{DM}(a)). (ii) The background was
measured using the same technique with $e^+$ identified instead of
$\mu^+$, relying on the large suppression of the $e^+$ rate
compared with the $\mu^+$ rate. (iv) The reaction $D^{*0}\to
\gamma D^0$; $D^0\to K^-\pi^+$, where the $\pi^+$ is first found
and then ignored was used to evaluate efficiencies. The published
result was
\begin{equation}
{{\Gamma(D_s^+\to\mu^+\nu)}\over{\Gamma(D_s^+\to\phi\pi^+)}}=0.173\pm
0.023 \pm 0.035 ~.
\end{equation}

%(M. Chadha et al. Phys. Rev. D58, 32002 (1998))

BaBar recently performed an improved analysis based on these
techniques \cite{BaBarDs}. They used 230 fb$^{-1}$ of continuum
data. To reduce the background and systematic errors they fully
reconstruct a $D^0$, $D^+$ or $D^*$ meson in the event with the
$\gamma$ and $\mu^+$ candidate. Their data are shown in
Fig.~\ref{DM}(b). They find

\begin{equation}
{{\Gamma\left(D_s^+\to\mu^+\nu\right)}\over{\Gamma\left(D_s^+\to\phi\pi^+\right)}}=0.143\pm
0.018 \pm 0.006 ~.
\end{equation}

%B. Aubert et al [hep-ex/0607094]

Both of these results, however, need to assume a value for ${\cal
{B}}(D_s^+\to\phi\pi^+)$ \cite{Yao:2006px}, in order to extract the
decay constant. Because of interferences among the final state
$K^+K^-\pi^+$ particles, the rate for $\phi\pi^+$ depends on
experimental cuts \cite{Stone-fpcp}, and thus has an inherent,
sizable,  systematic error. (Other experiments also normalize with
respect to this or other less well known modes.)

CLEO-c eliminates this uncertainty by making absolute measurements.
We tag a $D_s^-$ decay and search for three separate decay modes of
the $D_s^+$: (1) $\mu^+\nu$ and  $\tau^+\nu$, where (2)
$\tau^+\to\pi^+\overline{\nu}$ or (3) $\tau^+\to e^+\nu
\overline{\nu}$ \cite{Moscow}. For the first two analyses we require
the detection of the $\gamma$ from the $D_s^*\to\gamma D_s$ decay,
irrespective if the $D_s^*$ is the parent of the tag or the leptonic
decay. In either case, for real $D_s^*D_s$ events, the missing mass
squared recoiling against the photon and the $D_s^-$ tag should peak
at $M_{D_{s}^+}$ and is given by
\begin{equation}
{\rm MM}^{*2}=\left(E_{\rm CM}-E_D-E_{\gamma}\right)^2-
\left(\overrightarrow{p}_{\!\rm
CM}-\overrightarrow{p}_{\!D}-\overrightarrow{p}_{\!\gamma}\right)^2,\nonumber
\end{equation}
where $E_{\rm CM}$ ($\overrightarrow{p}_{\!\rm CM}$) is the center
of mass energy (momentum), $E_{D}$ ($\overrightarrow{p_D}$) and
$E_{\gamma}$ ($\overrightarrow{p_{\gamma}}$) are the energy of the
fully reconstructed $D_s^-$ tag, and the additional photon. In
performing this calculation we use a kinematic fit that constrains
the decay products of the $D_s^-$  to $M_{D_{s}^+}$ and conserves
overall momentum and energy.

\begin{figure}[htb]
\vspace{-3mm}
\includegraphics[width=0.3\textwidth]{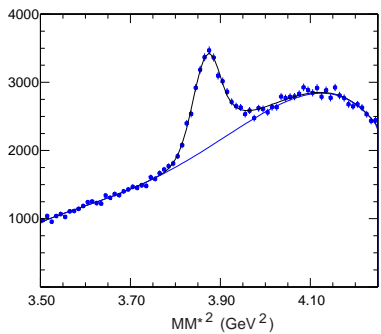}
 %\centerline{\epsfig{file=MMstar3.eps,width=2.9in}}
 \vspace{-2mm}
\caption{The MM*$^2$ distribution from events with a photon in
addition to the $D_s^-$ tag. The curve is a fit to the Crystal Ball
function and a 5th order Chebychev background function.}
\label{MMstar2}
\end{figure}

The MM$^{*2}$ from the $D_s^-$ tag sample data is shown in
Fig.~\ref{MMstar2}. There are 11880$\pm$399$\pm$511 signal events in
the interval $3.978>$MM$^{*2}>3.776$ GeV$^2$.

Candidate $D_s^+\to\mu^+\nu$ events are searched for by selecting
events with only a single extra track with opposite sign of charge
to the tag; we also require that there not be an extra neutral
energy cluster in excess of 300 MeV. Since here we are searching for
events where there is a single missing neutrino, the missing mass
squared, MM$^2$, evaluated by taking into account the seen $\mu^+$,
$D_s^-$, and the $\gamma$ should peak at zero, and is given by
\begin{eqnarray}
\label{eq:mm2} {\rm MM}^2&=&\left(E_{\rm
CM}-E_{D}-E_{\gamma}-E_{\mu}\right)^2\\\nonumber
         &&  -\left(\overrightarrow{p}_{\!\rm CM}-\overrightarrow{p}_
         {\!D}-\overrightarrow{p}_{\!\gamma}
           -\overrightarrow{p}_{\!\mu}\right)^2,
\end{eqnarray}
where $E_{\mu}$ ($\overrightarrow{p}_{\!\mu}$) is the energy
(momentum) of the candidate muon track.

We also make use of a set of kinematical constraints and fit the
MM$^2$ for each $\gamma$ candidate to two hypotheses one of which is
that the $D_s^-$ tag is the daughter of a $D_s^{*-}$ and the other
that the $D_s^{*+}$ decays into $\gamma D_s^+$, with the $D_s^+$
subsequently decaying into $\mu^+\nu$.

The kinematical constraints are the total momentum and energy, the
energy of the either the $D_s^*$ or the $D_s$, the appropriate
$D_s^* - D_s$ mass difference and the invariant mass of the $D_s$
tag decay products.
 This gives us a total of 7 constraints. The
missing neutrino four-vector needs to be determined, so we are left
with a three-constraint fit. We perform a standard iterative fit
minimizing $\chi^2$. As we do not want to be subject to systematic
uncertainties that depend on understanding the absolute scale of the
errors, we do not make a $\chi^2$ cut, but simply choose the photon
and the decay sequence in each event with the minimum $\chi^2$.

\begin{figure}[htbp]
 \vskip -0.2cm
%\centerline{ \epsfxsize=3.0in
\centerline{ \epsfxsize=2.4in \epsffile{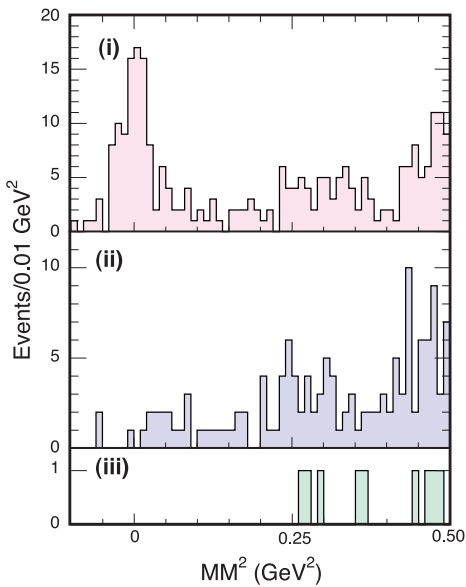}}
\vskip -0.2cm \caption{The MM$^2$ distributions from data using
$D_s^-$ tags and
 one additional opposite-sign
charged track and no extra energetic showers (see text).}
\label{mm2-data}
 \end{figure}

We consider three mutually exclusive cases: (i) the track deposits
$<$~300 MeV in the calorimeter, characteristic of a non-interacting
$\pi^+$ or a $\mu^+$; (ii) the track deposits $>$~300 MeV in the
calorimeter, characteristic of an interacting $\pi^+$; (iii) the
track satisfies our $e^+$ selection criteria. The MM$^2$
distributions are shown in Fig.~\ref{mm2-data}. The separation
between $\mu^+$ and $\pi^+$ is not unique. Case (i) contains 99\% of
the $\mu^+$ but also 60\% of the $\pi^+$, while case (ii) includes
1\% of the $\mu^+$ and 40\% of the $\pi^+$ \cite{DptomunPRD}.
There is a clear peak in Fig.~\ref{mm2-data}(i), due to
$D_s^+\to\mu^+\nu$. Furthermore, the events in the region between
$\mu^+\nu$ peak and 0.20 GeV$^2$ are dominantly due to the
$\tau^+\nu$, $\tau^+\to\pi^+\overline{\nu}$ decay. The best result
comes from summing case (i) and case (ii)  below MM$^2$ of 0.20
GeV$^2$; higher values of MM$^2$ admit background from $\eta\pi^+$
and $K^0\pi^+$ final states.
 The branching fractions are
summarized in Table~\ref{tab:results}. The absence of any detected
$e^+$ opposite to our tags allows us to set the upper limit listed
in Table~\ref{tab:results}.

\begin{table}
\caption{Measured $D_s^+$ Branching Fractions \label{tab:results}}
\begin{ruledtabular}
\begin{tabular}{lc}
    Final State  &  ${\cal{B}}$ (\%)       \\\hline
$\mu^+\nu$ & $0.657\pm 0.090\pm0.028$\\
$\mu^+\nu^{\dagger}$ & $0.664\pm 0.076\pm0.028$\\
$\tau^+\nu,~(\tau^+\to\pi^+\nu)$ & $7.1\pm 1.4\pm0.3$\\
$\tau^+\nu,~(\tau^+\to e^+\nu\bar{\nu})$ & $6.29\pm 0.78\pm0.52$\\
$\tau^+\nu$ (average) & $6.5\pm 0.8$\\
$e^+\nu$ & $< 3.1\times 10^{-4}$ (90\% cl)\\
\end{tabular}
\end{ruledtabular}
$\dagger$ From summing the  $\mu^+\nu$ and $\tau^+\nu$ contributions
for MM$^2$ $<$ 0.20 GeV$^2$.
\end{table}

\begin{figure}[htb]
%\centerline{ \epsfxsize=3.0in
%\centerline{ \epsfxsize=2.8in \epsffile{ecc-8.ps}}
\centerline{\psfig{file=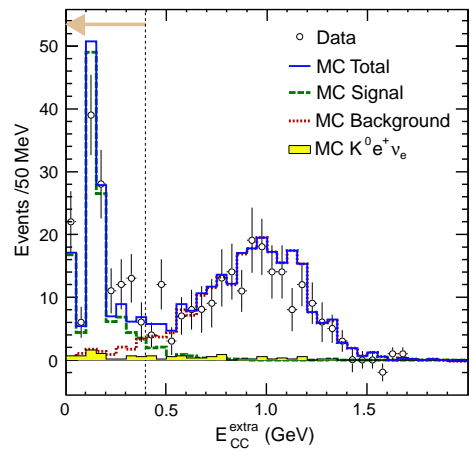,width=2.4in}} \vskip -2mm
 \caption{The extra calorimeter energy from data (points), compared with
 the Monte Carlo simulated estimates of semileptonic decays in general (dotted),
 the $K^0 e^+\nu$ mode specifically (shaded),
 as a sub-set of the semileptonics, and the expectation from signal (dashed).
The peak near 150 MeV is due to the $\gamma$ from $D_s^*\to\gamma
D_s$ decay. (The sum is also shown (line).) The arrow indicates the
selected signal region below 0.4 GeV.} \label{ecc-8}
 \end{figure}

CLEO-c also uses $D_s^+\to \tau^+\nu$,
 $\tau\to e^+\nu\bar{\nu}$.
Electrons of opposite sign to the tag are detected in events without
any additional charged tracks, and determining the unmatched energy
in the crystal calorimeter (${\rm E^{extra}_{CC}}$). This energy
distribution is shown in Fig.~\ref{ecc-8}. Requiring ${\rm
E^{extra}_{CC}<}$ 400 MeV, enhances the signal. The branching ratio
resulting from this analysis is also listed in
Table~\ref{tab:results}.

CLEO-c's published result for $f_{D^+}$ \cite{our-fDp} uses the
``double-tag" method at 3770 GeV, where $D^+D^-$ final states are
produced without any extra particles. Here one $D^-$ is fully
reconstructed and then there are enough kinematic constraints to
search for $D^+\to\mu^+\nu$ by constructing the missing mass-squared
(MM$^2$) opposite the $D^-$ and the muon. Fifty signal events are
found of which 2.8 are estimated background, resulting in:
\begin{equation}
{\cal{B}}(D^+\to\mu^+\nu)=(4.40\pm 0.66^{+0.09}_{-0.12})\times
10^{-4}~.
\end{equation}

The decay constant $f_{D^+}$ is obtained from Eq.~(\ref{eq:main})
using 1.040$\pm$0.007 ps as the $D^+$ lifetime, and $|V_{cd}|$ =
0.2238$\pm$0.0029, giving
\begin{equation}
f_{D^+}=(222.6\pm 16.7^{+2.8}_{-3.4})~{\rm MeV}~.
\end{equation}

CLEO-c also sets limits on ${\cal{B}}(D^+\to e^+\nu_{e})<2.4\times
10^{-5},$ \cite{our-fDp} and ${\cal{B}}(D^+\to\tau^+\nu)$ branching
ratio to $<2.1\times 10^{-3}$ at 90\% C.L. \cite{Rubin:2006nt}.
These limits are consistent with SM expectations.

For $D_s^+$ decays, we first test lepton universality in
\begin{equation}
R\equiv \frac{\Gamma(D_s^+\to \tau^+\nu)}{\Gamma(D_s^+\to
\mu^+\nu)}= 9.9\pm 1.9~, \label{eq:tntomu2}
\end{equation}
consistent with the predicted value of 9.72. Combining our branching
ratios determinations and using $\tau_{D_s^+}$=0.49 ps and
$|V_{cs}|$=0.9737, we find
\begin{eqnarray}
&& f_{D_s}=(280.1\pm 11.6 \pm 6.0) {~\rm MeV}, {\rm~and}\\
&&f_{D_s^+}/f_{D^+}=1.26\pm 0.11\pm 0.03.\nonumber
\end{eqnarray}

These preliminary results are consistent with most recent
theoretical models. As examples, unquenched lattice \cite{Lat:Milc}
predicts $1.24\pm0.01\pm0.07$, while one quenched lattice
calculation \cite{Lat:Taiwan} gives $1.13\pm0.03\pm0.05$, with other
groups having similar predictions \cite{others}.

\section{Semileptonic Decays}
One of the best ways to measure magnitudes of CKM elements is to use
semileptonic decays since they are far simpler to understand than
hadronic decays and the decay width is $\sim|V_{cq}|^2$. On the
other hand, measurements using other techniques have obtained useful
values for $V_{cs}$ and $V_{cd}$ \cite{artuso-CKM}, and thus
semileptonic $D$ decay measurements are a good laboratory for
testing theories of QCD. For a $D$ meson decaying into a single
hadron ($h$), the decay rate can be written exactly in terms of the
four-momentum transfer defined as:
\begin{equation}
q^2=\left(p^{\mu}_D-p^{\mu}_h\right)^2=m_D^2+m_h^2-2E_hm_D~.
\end{equation}

For decays to pseudoscalar mesons and ``virtually massless" leptons,
the decay width is given by:
\begin{equation}
{{d\Gamma(D\to P e^+\nu)}\over{dq^2}}
={{|V_{cq}|^2G_F^2p_P^3}\over{24\pi^3}}\left|f_+(q^2)\right|~,
\end{equation}
where $p_P$ is the three-momentum of $P$ in the $D$ rest frame,
and $f_+(q^2)$ is a ``form-factor," whose normalization must be
calculated theoretically, although its shape can be measured.

The shape measurements can distinguish between form-factor
parameterizations. In general,
\begin{equation}
f_+(q^2)=\frac{f_+(0)}{(1-\alpha_p)(1-\frac{q^2}{m^2_{\rm pole}})}
+\frac{1}{\pi}\int^{\infty}_{(M_D+M_P)^2}\!\!\!\!\!\!\!\!dq'^2\frac{Im
~f(q'^2)}{q'^2-q^2}, \nonumber
\end{equation}
which incorporates the possibility of a virtual of a nearby pole
(first term) with fractional strength $\alpha_p$. The integral term
can be expressed in terms of an infinite series \cite{Hill}.
Typically it takes only a few terms to describe the data. An
analytical parametrization
\begin{equation}
\label{eq:ff} f_+(q^2)=\frac{f_+(0)}{(1-{q^2}/{m^2_{\rm
pole}})(1-{\alpha q^2}/{m^2_{\rm pole}})}
\end{equation}
has become popular \cite{BK}, though it has been criticized as being
overly constraining \cite{Hill}. Fits are typically done for
$f_+(0)$ and either ${m_{\rm pole}}$ or $\alpha$. Naively, setting
$\alpha$ to zero gives the simple pole model where the pole mass
corresponds to the first vector resonance in the $D$-$P$ system,
$D_s^*$ for $D\to K e^-\nu$ and $D^*$ for $D\to \pi e^-\nu$

\begin{figure}[htb]
%\vskip 0.00cm \centerline{ \epsfxsize=3.0in
\centerline{
\includegraphics[width=8.5cm]{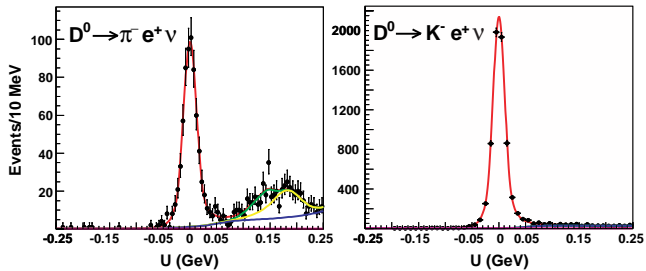}
} \vspace{-4mm}\caption{$U$ distributions using $D$ tags in
conjunction with an identified electron of opposite flavor plus a
single hadron. The peak centered at zero is signal. The dashed
curves indicate various backgrounds, while he solid curve shows the
fit to signal plus background.} \label{kpienu_U}
\end{figure}

CLEO-c uses two methods to analyze pseudoscalar decays. In the first
method tags are fully reconstructed and events with a missing $\nu$
are inferred using the variable $U=E_{\rm miss}-|P_{\rm miss}|$,
similar to MM$^2$, where ``miss" here refers to the missing energy
or momentum (see Fig.~\ref{kpienu_U}).

\begin{figure}[htb]
%\vskip 0.00cm \centerline{ \epsfxsize=3.0in
\centerline{
\includegraphics[width=8.5cm]{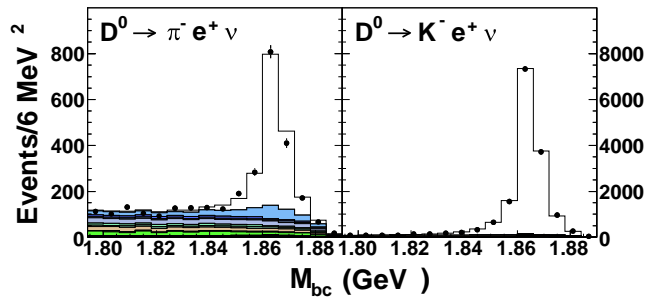}
}\vspace {-3mm} \caption{M$_{\rm bc}$ distributions for events
containing an identified electron plus a single hadron candidate.
The shaded regions indicate various backgrounds.}
\label{UntagKpienu_Mbc}
\end{figure}

The second method consists of also using missing energy and
momentum, skipping the step of reconstructing the tag, but using all
of the measured charged tracks and photons. Then the $D$ mass is
reconstructed. The beam-constrained mass (M$_{\rm bc}$)
distributions are shown in Fig.~\ref{UntagKpienu_Mbc}

Both cases have excellent signal to background in these modes. The
$\nu$-reconstruction has better statistical albeit poorer systematic
errors. Eventually combined results will be quoted; they should not
be averaged as there are a substantial number of events in common.

\begin{figure}[htb]
\vskip -0.3cm\centerline{
\includegraphics[width=4.5cm]{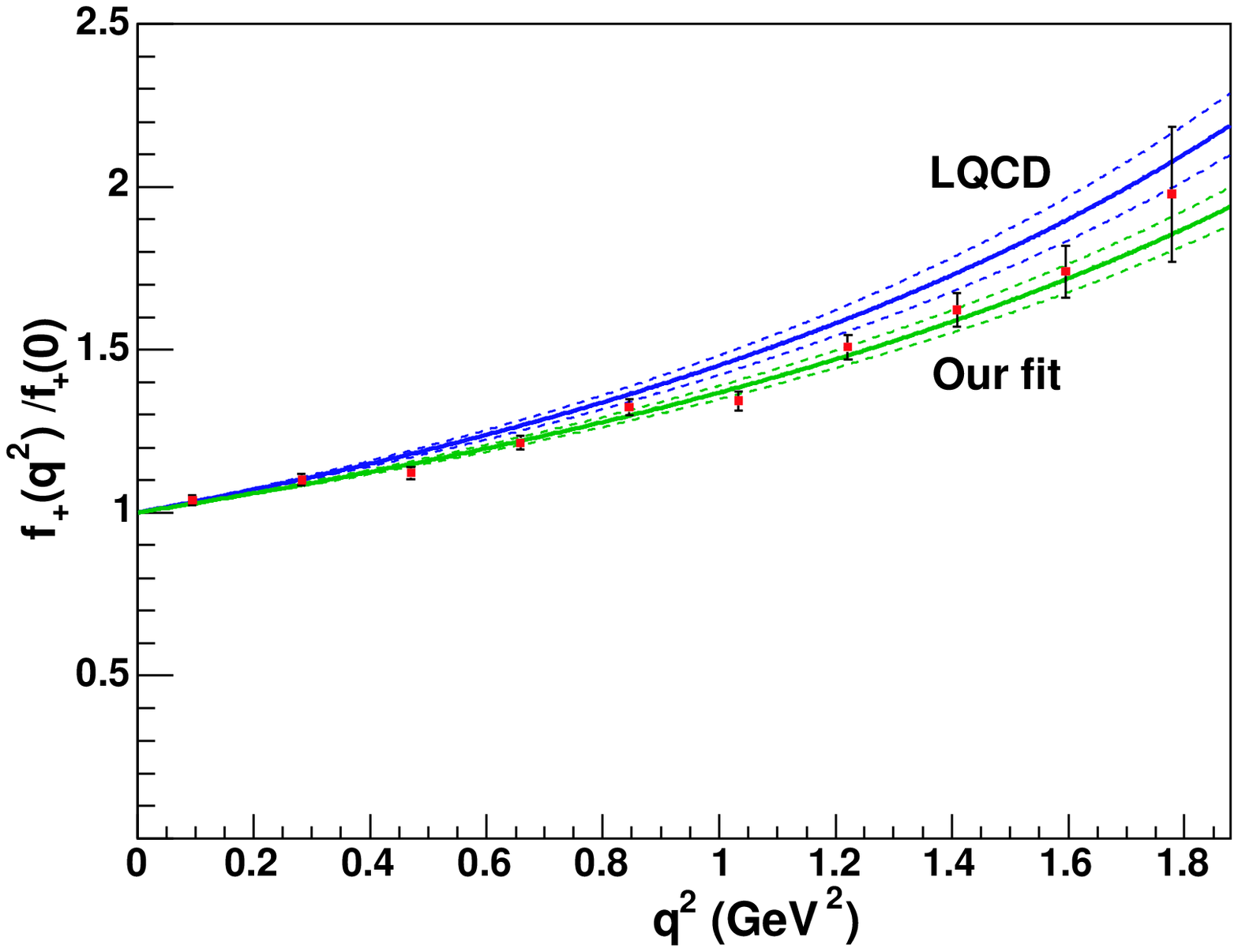}
\includegraphics[width=4.5cm]{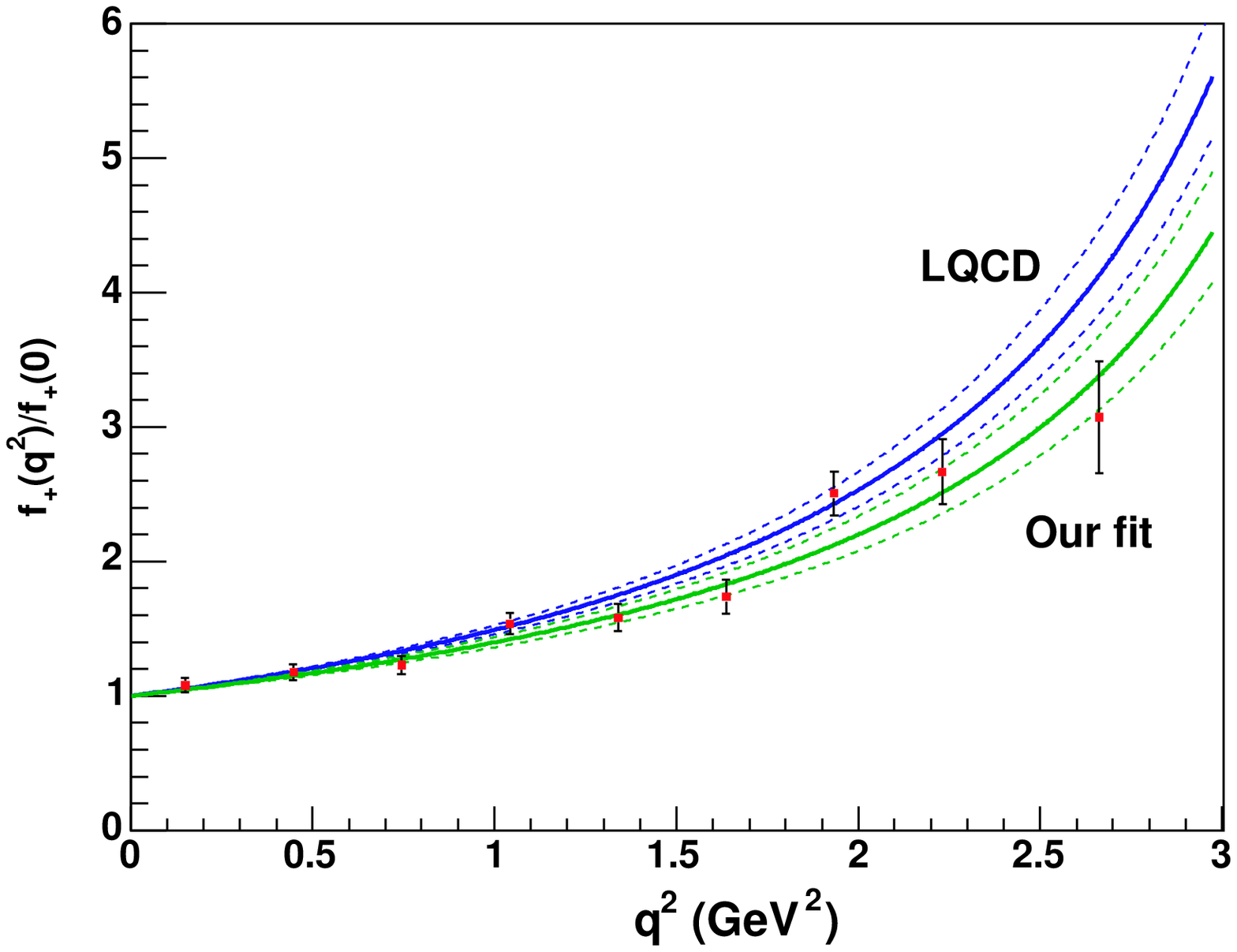} } \vspace{-2mm}\caption{CLEO-c form-factor
shapes using the tagged sample. The lower curves are fits to the
modified pole model, while the upper curves are fits to unquenched
lattice QCD \cite{Aubin}.} \label{forms}
\end{figure}

Form-factor shapes using the tagged sample are are shown in
Fig.~\ref{forms}.
The unquenched lattice QCD model \cite{Aubin} is systematically
higher than our data, but not in significant disagreement.
Properties of these decays are listed in Table~\ref{tab:DPenu}.

\begin{table}
\caption{Properties of $D^0\to P^- e^+\nu$ decays (preliminary)
\cite{Penu}. To determine $m_{\rm pole}$, $\alpha$ in
Eq.~\ref{eq:ff} is set to zero. \label{tab:DPenu}}
\begin{ruledtabular}
\begin{tabular}{lccl}
Quantity & $K^-e^+\nu$ & $\pi^- e^+\nu$& Source \\\hline
$\cal{B}$(\%) & 3.58(5)(5)&0.309(12)(6)& CLEO-c Tag\\
$\cal{B}$(\%) & 3.56(3)(11)&0.301(11)(10)& CLEO-c NoTag\\
$\cal{B}$(\%) & 3.58(18)&0.360(60)& PDG04\\
$|f_+(0)|$ & 0.761(10)(7) & 0.660(28)(11) & CLEO-c Tag\\
$|f_+(0)|$ & 0.749(5)(10) & 0.636(17)(13) & CLEO-c NoTag\\
$m_{\rm pole}$ (GeV) & 1.96(3)(1) & 1.95(4)(2) & CLEO-c Tag\\
$m_{\rm pole}$ (GeV) & 1.97(2)(1) & 1.89(3)(1) & CLEO-c NoTag\\
$\alpha$ & 0.22(5)(2) & 0.17(10)(5)& CLEO-c Tag\\
$\alpha$ & 0.21(4)(3) & 0.32(7)(3)& CLEO-c Tag\\
\end{tabular}
\end{ruledtabular}
\end{table}

\begin{figure}[htb]
%\vskip 0.00cm \centerline{ \epsfxsize=3.0in
\centerline{
\includegraphics[width=8.7cm]{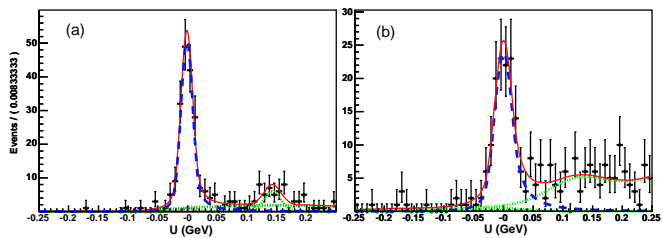}} \caption{$U$ distributions
for (a) $D^+$ and (b) $D^0$ decays into $\rho e^+\nu$ candidates.
The dashed curve shows the signal, the dotted curves show various
backgrounds and the solid curve the sum.} \label{rhoenuU}
\end{figure}

Measurements of the vector decays $D\to K^{*}e^+\nu$ and $\rho
e^+\nu$ can be used to determine $|V_{ub}|$ along with
measurements of $B\to\rho\ell^-\bar{\nu}$ and $B\to
K^*\ell^+\ell^-$ \cite{GP}. CLEO-c has examined $D$ vector
semileptonic decays. Non-parametric form-factors in the Cabibbo
favor $D^0\to K^{*0}e^+\nu$ decays have been measured by CLEO-c
\cite{CLEO-KS}, following a method developed by FOCUS
\cite{FOCUS-KS}. Cabibbo suppressed form-factors have been
measured in $D\to \rho e^+\nu$ decays. The $U$ distribution for
$\rho e^+\nu$ decays is shown in Fig.~\ref{rhoenuU}.
Preliminary branching fractions are listed in Table~\ref{tab:rare}
along with observations (or limits) from other rare semileptonic
decays. Selected candidates are used to measure the ratios of pole
dominated form-factor ratios as $R_V=1.40\pm 0.25 \pm 0.03$ and
$R_2=0.57\pm 0.18\pm 0.06$, using both charge and neutral modes
\cite{Penu}.

Other results on semileptonic decays from CLEO-c include measurement
of the inclusive $D^0$ and $D^+$ semileptonic branching fractions of
($6.46\pm 0.17\pm 0.13$)\%, and ($16.13\pm 0.20\pm 0.33$)\%,
respectively, leading to a measurement of the partial width ratio of
$\Gamma(D^+)/\Gamma(D^0)=(0.985\pm 0.028\pm 0.015)$, consistent with
isospin symmetry \cite{inclusive}.

\begin{table}
\caption{Rare semileptonic decay branching fractions
\label{tab:rare}}
\begin{ruledtabular}
\begin{tabular}{lc}
    Decay  &  ${\cal{B}} \times 10^{-4}$       \\\hline
$D^0\to\rho^-e^+\nu$ & $15.6\pm 1.6\pm 0.9$ \\
$D^+\to\rho^0e^+\nu$ & $23.2\pm 2.0\pm 1.2$ \\
$D^+\to \omega e^+\nu$ & $14.9\pm 2.7\pm 0.5$\\
$D^+\to \phi e^+\nu$ & $<2$ at 90\% CL\\
$D^+\to \eta e^+\nu$ & $12.9\pm 1.9\pm 0.7$\\
$D^+\to \eta' e^+\nu$ & $<3$ at 90\% CL\\
$D^0\to K^-\pi^+\pi^-e^+\nu$ & $2.9^{+1.9}_{-1.0}\pm 0.5$\\
\end{tabular}
\end{ruledtabular}
\end{table}

\section{Conclusions}

CLEO-c measurements of leptonic and semileptonic decays have
already reached precisions that provide very useful benchmarks for
testing of QCD theories. From leptonic decays we have
\begin{eqnarray}
f_{D^+}&=&(222.6\pm 16.7^{+2.8}_{-3.4})~{\rm MeV},\\\nonumber
f_{D_s^+}&=&(280.1\pm 11.6 \pm 6.0) {~\rm MeV},\\\nonumber
f_{D_s^+}/f_{D^+}&=&1.26\pm 0.11\pm 0.03~.
\end{eqnarray}
These results are consistent with most theoretical calculations
including those of unquenched lattice QCD \cite{others}.

CLEO-c is also breaking new ground in the study of semileptonic
decays. Form-factors in Cabibbo suppressed decays are reaching an
unprecedented level of accuracy and are also confronting theory.
\vspace{2mm}
\begin{acknowledgments}
I thank the U. S. National Science Foundation for support and my
CLEO colleagues for the excellent work that is reported here.  I had
useful conversations concerning this work with A. G. Akeroyd, M.
Artuso, H. Mahlke-Krueger, N. Menaa and C. S. Park.

\end{acknowledgments}


\begin{thebibliography}{99}
\bibitem{others}
See references to other theoretical predictions in \cite{our-fDp}.
\end{thebibliography}
\end{document}